# Asymmetrical Two-Level Scalar Quantizer with Extended Huffman Coding for Compression of Laplacian Source

Zoran Perić[1], Jelena Nikolić[1], Lazar Velimirović[2], Miomir Stanković[3], Danijela Aleksić[4]
[1]Faculty of Electronic Engineering, Univeristy of Niš, Aleksandra Medvedeva 14, 18000 Niš, Serbia
[2]Mathematical Institute of the Serbian Academy of Sciences and Arts, Kneza Mihaila 36, 11001 Belgrade, Serbia
[3]Faculty of Occupational Safety, University of Niš, Čarnojevića 10 A, 18000 Niš, Serbia
[4]Telekom Srbija, Voždova 11, 18000 Niš, Serbia
zoran.peric@elfak.ni.ac.rs, jelena.nikolic@elfak.ni.ac.rs, velimirovic.lazar@gmail.com,
miomir.stankovic@gmail.com, danijelaal@telekom.rs

*Abstract*—This paper proposes a novel model of the two-level scalar quantizer with extended Huffman coding. It is designed for the average bit rate to approach the source entropy as close as possible provided that the signal to quantization noise ratio (SQNR) value does not decrease more than 1 dB from the optimal SQNR value. Assuming the asymmetry of representation levels for the symmetric Laplacian probability density function, the unequal probabilities of representation levels are obtained, i.e. the proper basis for further implementation of lossless compression techniques is provided. In this paper, we are concerned with extended Huffman coding technique that provides the shortest length of codewords for blocks of two or more symbols. For the proposed quantizer with extended Huffman coding the convergence of the average bit rate to the source entropy is examined in the case of two to five symbol blocks. It is shown that the higher SQNR is achieved by the proposed asymmetrical quantizer with extended Huffman coding when compared with the symmetrical quantizers with extended Huffman coding having equal average bit rates.

## I. INTRODUCTION

The need for efficient data representation, a common interest in many practical digital communication systems, manifests the importance of signal compression in modern communication environments. Signal compression is usually categorized into quantization, as a lossy compression technique, and its lossless counterpart called entropy coding [1]–[5]. Both compression techniques have found wide application in various data representation needs. Entropy coding compresses data without loss of information but, in many cases, its achievable compression, itself bounded by the entropy of the source data, is insufficient for the purpose of low rate coding. By contrast, quantization can provide flexible compression for a wide range of bit rates at the cost of accordingly introduced quantization error or information loss. Therefore, it is important to research suitable lossy compression technique that provides the desired level of signal quality for the given bit rate. Lossless compression allows decreasing of bit rate without losing information and can be achieved using an entropy coding procedure [1]–[5]. There are many different types of entropy codes, the examples of which are Huffman, Golomb-Rice and arithmetic code [3]–[7]. In many modern applications, the combination of a quantizer and a lossless coder is used. Most often, due to simplicity, the quantizer and lossless coder are designed separately [4]–[7]. The obtained performances are not optimal, though. However, the desired performances can be obtained only with a joined design of the quantizer and lossless coder, which is done in this paper.

In this paper we propose a novel model of scalar quantizer with extended Huffman coding with a goal for its average bit rate to approach the source entropy as close as possible. The only constraint in the design of the novel quantizer is that the value of SQNR (Signal to Quantization Noise Ratio) decreases no more than 1 dB from the optimal SQNR Lloyd-Max's quantizer value. In fact, the two-level Lloyd-Max's quantizer [1]–[3] with zero decision threshold is a special case of our quantizer. Particularly, the novel two-level quantizer having non-negative variable decision threshold is designed depending on which SQNR has to be achieved. The basic idea described in this paper is that, unlike Lloyd-Max's quantizer, the asymmetry of representation levels is assumed to provide unequal probabilities of representation levels for the symmetric Laplacian probability density function (PDF). This in turn provides the proper basis for further implementation of lossless compression techniques. Among many lossless compression techniques, the most suitable one for utilization is extended Huffman coding technique that achieves the shortest average length of code words [3]–[5], [8], [9]. The analysis of extended Huffman code efficiency is provided in [9]. Furthermore, the performance analysis of four types of quantizers with Huffman coding for small and moderate bit rate are given in [10]. In the same paper, it is shown that the best performance is achieved by the hybrid quantizer composed of the uniform quantizer and Lloyd-Max's quantizer. The effective initialization problem of Lloyd-Max's quantizer's algorithm and the high design complexity of Lloyd-Max's quantizer with a large number of

This work is partially supported by Serbian Ministry of Education and Science through Mathematical Institute of Serbian Academy of Sciences and Arts (Project III44006) and by Serbian Ministry of Education and Science (Project TR32035).

quantization levels are pointed out in [11]. The lack of an effective implementation of Huffman coding technique on quantizers with a large number of quantization levels is shown in [5], [9], [10], [12]. For that reason, we propose a quantizer that has only two representation levels and we apply extended Huffman coding on the output levels of this quantizer. As with Lloyd-Max's quantizer, these representation levels are determined from the centroid condition. The design procedure of the asymmetrical scalar quantizer having the representation levels also determined in accordance with the centroid condition for the Laplacian and Gaussian source is given in [13] along with the analysis of the entropy when the distortion tends to be one.

This paper is organized as follows. Section II recalls some basic theory of scalar quantization and Lloyd-Max's quantizer. In addition, it gives the description of the design procedure for the proposed asymmetrical two-level scalar quantizer with variable decision threshold depending on SQNR. Section III gives a brief description to one of the most sophisticated and efficient lossless compression techniques, called extended Huffman coding technique. It also considers the application of extended Huffman coding on the output levels of the proposed asymmetrical two-level quantizer. The obtained numerical results are discussed in Section IV, and based on it, the conclusions about the possibilities of application of the proposed quantizer with extended Huffman coding are derived in Section V.

## II. DESIGN OF ASYMMETRICAL TWO-LEVEL SCALAR QUANTIZER WITH VARIABLE DECISION THRESHOLD DEPENDING ON SQNR

An $N$-level scalar quantizer $Q$ is defined by mapping $Q: R \rightarrow Y$ [1], [3], where $R$ is the set of real numbers, and:

$$Y \equiv (y_1, y_2, y_3, \ldots, y_N) \subset R \quad (1)$$

is a set of representation levels that makes the code book of size $|Y| = N$. Every $N$-level scalar quantizer partitions the set of real numbers into $N$ cells $R_i = (t_{i-1}, t_i]$, $i = 1, \ldots, N$, where $t_i$, $i = 0, 1, \ldots, N$ are decision thresholds and where it holds that $Q(x) = y_i$, $x \in R_i$. The quantizer designed iteratively in accordance with the centroid condition and the condition of the nearest neighbor is the optimal Lloyd-Max's quantizer [1]–[3]. The quantizer we propose in this paper is defined by the variable decision threshold along with the two representation levels determined from the centroid condition. Particularly, we determine this variable decision threshold depending on the quality, measured by SQNR that has to be achieved. Note that in the special case, when the mentioned variable decision threshold has zero value, the proposed quantizer becomes optimal. For the assumed Laplacian PDF of the unit variance [1]–[3]:

$$p(x) = \frac{1}{\sqrt{2}} \exp(-\sqrt{2}|x|) \quad (2)$$

the representation levels of the proposed quantizer are determined as follows:

$$y_1 = \frac{\int_{-\infty}^{t_1} xp(x)dx}{\int_{-\infty}^{t_1} p(x)dx} = \frac{\sqrt{2} + 2t_1}{2 - 4\exp(\sqrt{2}t_1)}, \quad (3)$$

$$y_2 = \frac{\int_{t_1}^{\infty} xp(x)dx}{\int_{t_1}^{\infty} p(x)dx} = t_1 + \frac{1}{\sqrt{2}}, \quad (4)$$

where the variable decision threshold is denoted by $t_1$. From the last two equations, it is obvious that the representation levels of the proposed quantizer are not symmetrical.

The performances of the quantizer are often determined by SQNR which is defined as follows [1]–[3]:

$$\text{SQNR} = 10 \log\left(\frac{\sigma^2}{D}\right), \quad (5)$$

and expressed in dB where $\sigma^2$ is the variance of the input signal $x$, while $D$ is the distortion added with quatization. Assuming the unit variance for the given range of SQNR values, one can firstly determine the appropriate $D$ values:

$$D = \frac{\sigma^2}{10^{\frac{\text{SQNR}}{10}}} = \frac{1}{10^{\frac{\text{SQNR}}{10}}}. \quad (6)$$

By further defining the distortion of the proposed quantizer:

$$D = \int_{-\infty}^{t_1} (x - y_1)^2 p(x)dx + \int_{t_1}^{\infty} (x - y_2)^2 p(x)dx, \quad (7)$$

and by combining with (3) and (4), we can derive a closed form expression for the distortion of the proposed quantizer as a function of the variable decision threshold $t_1$:

$$D = \frac{3 - 4\exp(\sqrt{2}t_1) + 2\sqrt{2}t_1 + 2t_1^2}{2 - 4\exp(\sqrt{2}t_1)}. \quad (8)$$

Using this expression, one can find the corresponding threshold value for the given distortion value, hence, the design of the proposed quantizer is enabled.

## III. APPLICATION OF EXTENDED HUFFMAN CODING ON ASYMMETRICAL TWO-LEVEL SCALAR QUANTIZER

Output levels of a quantizer can be considered as a discrete source of symbols and can be coded using fixed-length codewords. However, a more effective manner of coding is by using entropy code with variable-length codewords [1]–[5], [14], [15]. The bit rate of any lossless code is always higher than the entropy, where the aim is to approach the entropy as close as possible. To achieve this, symbols with large probabilities are coded with short codewords and less-probable symbols are coded with longer codewords. As aforementioned, there are many types of entropy codes, such as Huffman code, arithmetic code and Golomb-Rice code [3]–[7]. In this section we consider the application of extended Huffman coding on the asymmetrical two-level quantizer defined in the previous section. The procedure of Huffman coding includes determining the optimal length of code words for a given probability of symbols [3]–[5], [8]. It is sometimes beneficial to additionally reduce the bit rate by blocking more than one symbol together. In the mentioned cases, extended Huffman coding technique is used. Particularly, extended Huffman coding is the procedure of determining the optimal length of code words for blocks of two or more symbols [3], [4], [8], [9].

Let us denote by $p_1$ the probability that a sample of the input signal has a lower value than the value of decision

threshold $t_1$:

$$p_1 = \int_{-\infty}^{t_1} p(x)dx = 1 - \frac{1}{2}\exp(-\sqrt{2}t_1), \quad (9)$$

and by $p_2$, the probability that a sample of the input signal has a greater value than the value of decision threshold $t_1$:

$$p_2 = \int_{t_1}^{\infty} p(x)dx = \frac{1}{2}\exp(-\sqrt{2}t_1). \quad (10)$$

These probabilities actually refer to the symbol probabilities, i.e. to the probabilities of the occurence of representation levels $y_1$ and $y_2$. Since we consider a two-level quantizer, in fact, we observe a two symbol source. We can now define the probabilities of two to more symbol blocks as:

$$P_{i,j} = p_i p_j, \quad i=1,2, \; j=1,2, \quad (11)$$

$$P_{i,j,\ldots,k} = p_i p_j \ldots p_k, \; i=1,2, \; j=1,2,\ldots,k=1,2. \quad (12)$$

Note that blocking two symbols together means that the symbol alphabet size goes from $M$ to $2^M$, where $M$ is the size of the initial symbol alphabet. In this paper, we consider four cases, of two, three, four and five symbol blocks, so that the size of the extended alphabet is 4, 8, 16 and 32, respectively. For the proposed quantizer with extended Huffman coding we examine the convergence of the average bit rate to the source entropy. The source entropies for two and more symbol blocks are given by the following expressions, respectively [4]:

$$H = \sum_{i=1}^{2}\sum_{j=1}^{2} P_{i,j} \, ld \frac{1}{P_{i,j}}, \quad (13)$$

$$H = \sum_{i=1}^{2}\sum_{j=1}^{2}\ldots\sum_{k=1}^{2} P_{i,j,\ldots,k} \, ld \frac{1}{P_{i,j,\ldots,k}}. \quad (14)$$

The average bit rate of the observed quantizer in the case of two symbol blocks can be determined as:

$$\overline{R} = \sum_{i=1}^{2}\sum_{j=1}^{2} P_{i,j} l_{i,j}, \quad (15)$$

where $l_{i,j}$, $i = 1, 2, j = 1, 2$ stand for the length of the code words. Similarly, the average bit rate in the case of more symbol blocks is determined by:

$$\overline{R} = \sum_{i=1}^{2}\sum_{j=1}^{2}\ldots\sum_{k=1}^{2} P_{i,j,\ldots,k} l_{i,j,\ldots,k}. \quad (16)$$

The procedure of determining the length of the code words using extended Huffman coding and the code book construction includes the following steps:

**Step 1.** Determining the symbol block probabilities and sorting them in the descending order. Assigning appropriate probabilities to the initial nodes of the graph.

**Step 2.** Application of an iterative process. In each iteration the two nodes with the smallest probabilities are connected and the sum of their probabilities is assigned to a new node. Processing further until the nodes' sum of the probabilities joining in the last step equals one.

**Step 3.** The construction of code words. The code word for each symbol is determined by beginning from the tree root (node with probability 1) and branches, to which the allocation of zero value (upper branch) and 1 (lower branch) is acquired. The assignment process continues to the left until all possible branches are covered. The code word is formed from zeros and ones that are on the path from the root to the node corresponding to that symbol.

## IV. NUMERICAL RESULTS AND DISCUSSION

Numerical results presented in this section for the proposed quantizer with extended Huffman coding are obtained for the cases where the SQNR value does not decrease more than 1 dB from the optimal quantizer SQNR value. The optimal SQNR value of the Lloyd-Max's quantizer having two quantization levels is 3 dB. Therefore, the SQNR range in which we consider the performance of the proposed quantizer is from 2 dB to 3 dB. The calculated performance of the proposed quantizer with extended Huffman coding in the case of two, three, four and five symbol blocks are shown in Fig. 1 and Table I. From the shown numerical results one can notice that the average bit rate of the proposed quantizer with extended Huffman coding approaches the source entropy where this convergence is greater in the case of five symbol blocks than in the other observed cases. However, by blocking more and more symbols together, extended Huffman coding technique becomes impractical since the complexity of extended Huffman coder increases as well [4], where the decrease of the average bit rate is not significant (see Table I). Accordingly, in this paper our analysis is mainly constrained to the case of three symbol blocks. From the results given in Table I and Fig. 1 one can observe that when the SQNR value decreases up to 0.5 dB from the optimal SQNR value, there is a little deviation of the average bit rate from the source entropy in the case of three symbol blocks.

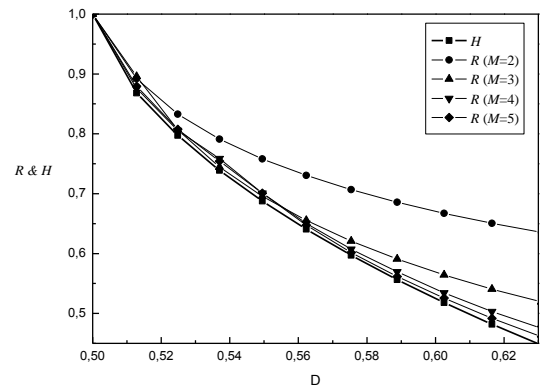

Figure 1. The dependency of the average bit rate and the source entropy on the distortion for the proposed quantizer with extended Huffman coding

However, when the deviation of SQNR is in the range of 0.5 dB to 1 dB, a slightly larger deviation of the average bit rate from the source entropy can be perceived. Observe that in both ranges the average bit rate and the source entropy converge more closely in the case of three symbol blocks than in the case of two symbol blocks. It is important to notice that for the proposed quantizer with extended Huffman coding in the case of three symbol blocks with an average bit rate reduction of 0.35 bits, the reduction in SQNR of 0.5 dB is achieved. This is about 0.9 dB smaller SQNR reduction for the same amount of the compression than the one ascertained in the considered range of the average bit rate [4]. This result is not achievable by any other known procedure of designing symmetrical quantizers. Hence, the advantage of our asymmetrical quantizer over the symmetrical one is that the higher signal quality, measured by SQNR, can be achieved for the same average bit rate. Accordingly, it is obvious that the proposed quantizer stands for a very efficient coding solution. Finally, from the last row in Table I one can notice that optimal Lloyd-Max's quantizer is actually the special case of the proposed quantizer Particularly, when the decision threshold $t_1$ of the proposed quantizer is settled to zero, the proposed quantizer is Lloyd-Max's quantizer that has the symmetrical representational levels, i.e. equal probabilities $p_1$ and $p_2$. For such values of probabilities, the values of the entropy and the average bit rate of the proposed quantizer are equal and amount to one.

TABLE I. Performance of the proposed quantizer with extended Huffman coding in the case of two, three, four and five symbol blocks

| SQNR | D | $t_1$ | $p_1$ | $p_2$ | H | $\bar{R}$ (M=2) | $\bar{R}$ (M=3) | $\bar{R}$ (M=4) | $\bar{R}$ (M=5) |
|---|---|---|---|---|---|---|---|---|---|
| 2 | 0.6309 | 1.1876 | 0.9067 | 0.0932 | 0.4471 | 0.6353 | 0.5191 | 0.4749 | 0.4612 |
| 2.1 | 0.6165 | 1.1096 | 0.8958 | 0.1041 | 0.4819 | 0.6506 | 0.5406 | 0.5030 | 0.4918 |
| 2.2 | 0.6025 | 1.0324 | 0.8838 | 0.1161 | 0.5181 | 0.6673 | 0.5643 | 0.5343 | 0.5256 |
| 2.3 | 0.5888 | 0.9546 | 0.8703 | 0.1296 | 0.5564 | 0.6859 | 0.5909 | 0.5697 | 0.5616 |
| 2.4 | 0.5754 | 0.8756 | 0.8550 | 0.1449 | 0.5970 | 0.7067 | 0.6209 | 0.6073 | 0.6017 |
| 2.5 | 0.5623 | 0.7943 | 0.8373 | 0.1626 | 0.6405 | 0.7306 | 0.6555 | 0.6504 | 0.6475 |
| 2.6 | 0.5495 | 0.7091 | 0.8165 | 0.1834 | 0.6878 | 0.7582 | 0.6958 | 0.7008 | 0.7009 |
| 2.7 | 0.5370 | 0.6176 | 0.7912 | 0.2087 | 0.7390 | 0.7911 | 0.7445 | 0.7588 | 0.7542 |
| 2.8 | 0.5248 | 0.5147 | 0.7585 | 0.2414 | 0.7974 | 0.8328 | 0.8066 | 0.8056 | 0.8075 |
| 2.9 | 0.5128 | 0.3866 | 0.7105 | 0.2894 | 0.8680 | 0.8921 | 0.8958 | 0.8758 | 0.8796 |
| 3 | 0.5 | 0 | 0.5 | 0.5 | 1 | 1 | 1 | 1 | 1 |

## V. Conclusion

In this paper we have presented a novel class of asymmetrical quantizers with extended Huffman coding that are designed to provide the required quality of the quantized signal, measured by SQNR, and for the average bit rate to approach the source entropy as close as possible. Based on the performance analysis of the proposed quantizer with extended Huffman coding, one can conclude that the average bit rate and the source entropy converge more closely by blocking more symbols together. However, since the complexity of extended Huffman coder increases as well, the analysis presented in this paper is constrained to the case of three symbol blocks. Since it has been demonstrated that the proposed asymmetrical quantizer with extended Huffman coding stands for a very simple and efficient coding solution, better then the one based on symmetrical quantizer with extended Huffman coding, one can believe that it will find its way toward the practical implementation in signal compression.